\documentclass[a4paper,aps,pre,floatfix,twocolumn,showpacs,nofootinbib,superscriptaddress]{revtex4-1}
\usepackage{amssymb}
\usepackage{amsmath}
\usepackage{amsfonts}
\usepackage{epsfig}
\usepackage{graphicx}
\usepackage{color}
\usepackage[latin1]{inputenc}
\usepackage{hyperref}

\newcommand{\av}[1]{\langle {#1} \rangle}
\newcommand{\Ct}{C_{\mathrm{th}}}
\newcommand{\ct}{c_{\mathrm{th}}}
\newcommand{\MB}{Bogu\~{n}\'{a} \textit{et al.}}

\newcommand{\FigPath}{.}

\begin{document}

\title{Lifespan method as a tool to study criticality in
  absorbing-state phase transitions}

\author{Ang\'{e}lica S. Mata}
\affiliation{Departamento de F\'{\i}sica, Universidade Federal de
  Vi\c{c}osa, 36571-000, Vi\c{c}osa - MG, Brazil}

\affiliation{Departament de F\'\i sica i Enginyeria Nuclear,
  Universitat Polit\`ecnica de Catalunya, Campus Nord B4, 08034
  Barcelona, Spain}

\author{Marian Bogu\~{n}\'{a}} 

\affiliation{Departament de F\'{\i}sica Fonamental, Universitat de
  Barcelona, Mart\'{\i} i Franqu\`{e}s 1, 08028 Barcelona, Spain}

\author{Claudio Castellano} 

\affiliation{Istituto dei Sistemi Complessi (ISC-CNR), via dei Taurini
  19, I-00185 Roma, Italy}

\affiliation{Dipartimento di Fisica, ``Sapienza'' Universit\`a di
  Roma, P.le A. Moro 2, I-00185 Roma, Italy}

\author{Romualdo Pastor-Satorras} 

\affiliation{Departament de F\'\i sica i Enginyeria Nuclear,
  Universitat Polit\`ecnica de Catalunya, Campus Nord B4, 08034
  Barcelona, Spain}


\date{\today}

\begin{abstract}
  In a recent work, a new numerical method (the lifespan method) has
  been introduced to study the critical properties of epidemic processes
  on complex networks [Phys. Rev. Lett. \textbf{111}, 068701
  (2013)]. Here, we present a detailed analysis of the viability of this
  method for the study of the critical properties of generic
  absorbing-state phase transitions in lattices. Focusing on the well
  understood case of the contact process, we develop a finite-size
  scaling theory to measure the critical point and its associated
  critical exponents. We show the validity of the method by studying
  numerically the contact process on a one-dimensional lattice and
  comparing the findings of the lifespan method with the standard
  quasi-stationary method. We find that the lifespan method gives
  results that are perfectly compatible with those of quasi-stationary
  simulations and with analytical results.  Our observations confirm
  that the lifespan method is a fully legitimate tool for the study of
  the critical properties of absorbing phase transitions in regular
  lattices.
\end{abstract}

\maketitle

\section{Introduction}

A key class of dynamical non-equilibrium systems are those with
absorbing states, {\it i.e.}, states from which the dynamics cannot
escape once it falls on them. Classical examples of such systems are
epidemic spreading processes \cite{epidemics}; obviously, a fully
healthy state is absorbing in the above sense, provided we do not allow
for immigration of infected individuals. A very relevant feature of many
systems with absorbing states is their ability to exhibit
\textit{absorbing-state phase transitions} \cite{Henkel,Marrobook}, that
is, non-equilibrium phase transitions between an active state,
characterized by everlasting activity in the thermodynamic limit, and an
absorbing state, where activity is absent.

Apart from a few exactly solvable models \cite{Schutz20011}, the
theoretical characterization of absorbing-state phase transitions is
based usually on mean-field theories \cite{Marrobook}, field theory
renormalization procedures \cite{tauber2014}, topological phase-space
field theories \cite{PhysRevE.74.041101}, dynamical mean-field plus
coherent anomaly extrapolation \cite{PhysRevE.51.6261} or series
expansions for particular models
\cite{0305-4470-29-22-007,0305-4470-32-28-304}.  While simple mean-field
theory is only valid above the upper critical dimension, application of
other techniques in physical dimensions is usually hindered by technical
difficulties. For this reason, a large amount of our knowledge about the
properties of absorbing-state phase transitions is based on computer
simulation of different representative models. The numerical analysis of
this computer data represents a different sort of challenge, which is
mainly hampered by finite size effects. In finite systems, any
realization of the dynamics is bound to reach sooner or later the
absorbing state, even in the active phase, due to dynamic
fluctuations. This difficulty can be overcome by applying the
finite-size scaling technique \cite{cardy88}, based on the size
dependence of physical observables that are averaged only over surviving
runs, i.e., realizations which have not yet fallen into the absorbing
state \cite{Marrobook}.  The critical point and various critical
exponents can then be determined by studying the decay of the average of
different observables over surviving runs as a function of the system
size. Averaging over surviving runs is, however, computationally highly
inefficient. A more effective alternative is provided by the
quasi-stationary (QS) method
\cite{DeOliveira05,PhysRevE.73.036131,DickmanJPA}, in which every time
the system tries to visit an absorbing state, it jumps instead to an
active configuration previously stored during the simulation.

Recently, in the context of epidemic modeling on complex networks
\cite{Pastor-Satorras:2014aa}, \MB~\cite{PhysRevLett.111.068701},
building on the traditional method of seed simulations \cite{Henkel},
proposed to consider the lifespan of spreading simulations starting from
a single infected site as a tool to determine the position of the
critical point. Inspired by the satisfactory performance of the lifespan
method (LS) on epidemic models in networks
\cite{PhysRevLett.111.068701}, in this paper we consider its extension
and application to  models with absorbing states on regular
Euclidean lattices, presenting a detailed finite-size scaling theory for
this new approach to determine critical properties of absorbing state
phase transitions. To provide a concrete application example, we focus
on the well-known controlled case of the contact process (CP)
\cite{harris74} in a one dimensional lattice, for which theoretical and
high-quality numerical results are already available.  In this way, we
are able to make a direct assessment of the reliability of the LS
method.  A close comparison with the results of quasi-stationary
simulations is also performed.  Our results indicate that the lifespan
method is a perfectly viable alternative to investigate critical
properties of absorbing phase transitions in regular lattices.

We have organized our paper as follows: In
Sec.~\ref{sec:contact-process} we define the CP and present the
numerical implementation and main properties of this
model. Section~\ref{sec:quasi-stat-meth} reviews briefly the QS method
and the finite-size scaling form of the properties computed from
it. Sections~\ref{sec:lifespan-ls-methods} and~\ref{sec:fss} present the
LS method and discuss its finite-size scaling theory, respectively. In
Sec.~\ref{sec:1d}, we present numerical results comparing the
predictions of both QS and LS methods for the CP in a $d=1$
lattice. Conclusions and perspectives are finally discussed in
Sec.~\ref{sec:conclusions}.

\section{The contact process}
\label{sec:contact-process}

The contact process (CP) represents the simplest theoretical model with
an absorbing-state phase transition~\cite{harris74}. The CP is defined
as follows: Sites in a lattice are characterized by a binary variable
$\sigma_i$ that can take values $\sigma_i = 1$ (occupied by a particle)
or $\sigma_i = 0$ (empty). Each occupied vertex can spontaneously become
empty at a rate which, without loss of generality, is set equal to $1$,
thus fixing the time scale.  On the other hand, at a rate $\lambda / z$,
where $z$ is the coordination number of the lattice, an occupied site
creates offspring particles on its empty nearest neighbors (note that all temporal processes are assumed to be Poisson point processes). The creation
of particles is a catalytic process occurring exclusively in pairs of
empty-occupied sites, implying that the state devoid of particles is a
fixed point of the dynamics (i.e. an absorbing state).

On a lattice with $N$ nodes, the CP is numerically simulated as follows
\cite{Marrobook}: An occupied site $j$ is randomly selected.  With
probability $p=1/(1+\lambda)$ the selected site becomes empty.  With
complementary probability $1-p$ one of the neighbors of $j$ is randomly
chosen and, if empty, it becomes occupied.  Time is incremented by
$\Delta t=1/[(1+\lambda)n(t)]$, where $n(t)$ is the number of occupied
sites at time $t$. We note that this prescription \cite{Marrobook}
(consistent with a variation of the classical Gillespie
algorithm~\cite{GILLESPIE:1977db,GILLESPIE:1976rw} in which time is
incremented in a deterministic way) corresponds to a sequential update
of events. This is the only way to reproduce offspring creation events
among occupied and empty sites taking place at rate $\lambda/z$
according to a Poisson point process.

In an infinite system, the CP displays an absorbing-state phase
transition at a critical point $\lambda_c$, between an absorbing phase
for $\lambda \leq \lambda_c$, and an active one for $\lambda > \lambda_c$. The
order parameter of the transition is the stationary density of occupied
sites $\rho_{st}(\lambda) \equiv \lim_{t\rightarrow
  \infty}\lim_{N\rightarrow \infty} \langle n(t) \rangle/N$, which is
zero below the threshold $\lambda_c$ and larger than zero above it.
Near the critical point $\rho_{st}(\lambda)$ vanishes as a power law
\begin{equation}
  \rho_{st}(\lambda) \varpropto (\lambda - \lambda_c)^\beta,
  \label{eq:rho}
\end{equation}
characterized by the critical exponent $\beta$. The onset of critical
fluctuations at the transition is ruled by a diverging 
correlation length  $\xi$, given by
\begin{equation}
  \xi \varpropto |\lambda - \lambda_c|^{-\nu_{\bot}},
  \label{eq:cor}
\end{equation}
where $\nu_{\bot}$ is the finite size scaling exponent.

\section{The quasi-stationary  method}
\label{sec:quasi-stat-meth}

The standard numerical procedure to investigate the finite-size scaling
at absorbing phase transitions---by measuring the average of the order
parameter restricted only to surviving runs---is extremely inefficient,
since surviving configurations are very rare at long times. The
quasi-stationary method represents an alternative strategy which
consists in constraining the system to be in a quasi-stationary
state. In practice, this is implemented by replacing the absorbing
state, every time the system tries to visit it, with an active
configuration randomly taken from the history of the simulation
\cite{DeOliveira05}. For this task, a list of $M$ active configurations
is stored and constantly updated. An update consists in randomly
choosing a configuration in the list and replacing it by the present
active configuration with a small probability $p_r \Delta t \ll 1$.
The parameter $p_r$ is typically chosen to be equal to $0.02$. In any case, in the
simulations presented here, no significant dependence on this parameter
was detected for a wide range of variations in simulations.

After a relaxation time, the QS quantities are
determined during a given averaging time window. Following this
approach, it is possible to evaluate the full probability distribution
of the number of occupied vertices in the quasi-stationary state and
use it to calculate all quantities of interest.  The transition point
is then determined by considering the modified
susceptibility~\cite{Ferreira12}
\begin{equation}
  \chi =
  \frac{L^d (\langle\rho^2\rangle-\langle\rho\rangle^2)}
{\langle\rho\rangle}.
  \label{eq:chi}
\end{equation}
Close to the critical point, the susceptibility diverges as $\chi \sim
(\lambda_c - \lambda)^{-(\gamma+\beta)}$. As we see, the critical
exponent of this susceptibility is larger than the standard one
($\gamma$), which simplifies its numerical evaluation while preserving
all the scaling properties. In a finite lattice of side $L$, $\chi$
shows a diverging peak at $\lambda=\lambda_p^{QS}(L)$, providing a
finite size approximation of the critical point. In the thermodynamic
limit, $\lambda_p^{QS}(L)$ approaches the true critical point with the
scaling form~\cite{binder2010monte}
\begin{equation}
  \lambda_p^{QS}(L) = \lambda_c + A_{QS} L^{-1/\nu_\perp}.
  \label{eq:3}
\end{equation}
In finite but large systems, the density of occupied sites and the
susceptibility can be written near the critical point with the
finite-size scaling form\footnote{Here and in the following we will not
  consider the possibility of the breakdown of standard finite-size
  scaling forms due to dangerously irrelevant scaling fields
  \cite{ffs_breakdown}.}~\cite{Marrobook}
\begin{equation}
\rho_{st}(\lambda,L) \varpropto L^{-\beta/\nu_{\bot}}
f\left[(\lambda-\lambda_c) L^{1/\nu_{\bot}}\right], 
\label{eq:rho2}
\end{equation}
and
\begin{equation}
\chi(\lambda,L) \varpropto L^{(\gamma+\beta)/\nu_{\bot}}
g\left[(\lambda-\lambda_c) L^{1/\nu_{\bot}}\right], 
\label{eq:sus_standard}
\end{equation}
where $f(x)$ and $g(x)$ are scaling functions that satisfy
$f(x) \varpropto x^{\beta}$ for $x \gg 1$,
$f(x) \varpropto |x|^{-\nu_{\bot}+\beta}$ for $-x \gg 1$, and $f(x)= $
const. for $|x|\ll 1$, and $g(x) \varpropto |x|^{-(\gamma+\beta)}$ for
$|x| \gg 1$, $g(x)= $ const. for $|x| \ll 1$.  Equations~\eqref{eq:rho2}
and \eqref{eq:sus_standard} imply that, at the critical point, the QS
observables depend on $L$ as
\begin{equation}
\rho_{st}(\lambda_c,L) \varpropto 
  L^{-\beta/\nu_{\bot}}  \;\; \mbox{ and } \;\;  \chi(\lambda_c,L) \varpropto  L^{(\gamma+\beta)/\nu_{\bot}}.
\label{eq:FFS}
\end{equation}

\section{The lifespan method}
\label{sec:lifespan-ls-methods}

The LS method proposed by \MB~\cite{PhysRevLett.111.068701} considers
spreading simulations starting from a single occupied site.  Each
realization of the dynamical process is characterized by its lifespan
$\tau$ and its coverage $C$, where the latter is defined as the
number of distinct sites which have been occupied at least once during the
realization.  In the thermodynamic limit, realizations can be either
\textit{finite} or \textit{endemic}. Endemic realizations have an
infinite lifespan and their coverage is equal to the system size; 
such realizations
are only possible above the critical point. Finite realizations, on
the other hand, have finite lifespan and coverage. Finite realizations can
be found both below and above the critical point, although the
probability to find a finite realization decreases when $\lambda$ is
increased above the critical point.

In the LS method, the role of the order parameter is played by the
probability that a run is endemic, $\mathrm{P_{end}}(\lambda)$. This
probability is zero below the critical point and grows monotonously for
$\lambda >\lambda_c$, approaching $1$ in the limit $\lambda \rightarrow
\infty$. The role of the susceptibility is played by the average
lifetime of finite realizations $\av{\tau}$.  For small values of
$\lambda$ all realizations are finite and have a very short duration. As
$\lambda$ grows the average duration of finite realizations increases,
diverging at the critical point. Above the critical point, the
probability of a realization to be endemic increases and those
realizations that remain finite have necessarily a short lifespan. This
is so because once a realization has been alive for a very long time,
the probability that it becomes finally endemic increases. As a result,
$\av{\tau}$ diverges when approaching the critical point from the left
and decreases as $\lambda$ is increased further. In a finite system with $N$
nodes, $\av{\tau}$ exhibits a peak for a value $\lambda_p^{LS}(N)$ that
converges to $\lambda_c$ in the thermodynamic limit. 

In finite systems,
the program described above has to be implemented with care. Indeed, in
a finite system any realization is bound to end, reaching the absorbing
state, even though this might occur over astronomically long temporal
scales. Therefore, the distinction between finite and endemic
realizations is, a priori, not clear-cut. In practice, we declare a
realization as endemic whenever its coverage fraction reaches a predefined
threshold value $\Ct=\ct N$, with $\ct$ a constant value between zero and one. 
Realizations ending before the value $C=\Ct$ is
reached are considered to be finite. In the thermodynamic limit,
reaching $\Ct$ means that an infinite number of nodes have
been reached by the outbreak. If so, the probability that such
realization is eventually trapped in the absorbing state is zero,
meaning that the realization is endemic with probability 1.

\section{Finite-size scaling of the lifespan method}
\label{sec:fss}

In this section, we present a finite-size scaling theory of the LS
method, which enables the detailed analysis of numerical simulations. In
general, the theory can be applied to any type of discrete
structure. For this reason, hereafter we use the number of sites $N$ as
the measure of the size of the system. The case of a lattice of side $L$
in $d$ dimensions can be easily recovered by replacing $N=L^d$. Let
$\Psi(\tau,C;\lambda)$ be the joint probability of a realization of the
CP process to have, in an infinite size system, a (finite) lifespan
$\tau$ and coverage $C$. This joint probability can be written as
\begin{equation}
\Psi(\tau,C;\lambda) = \psi(\tau;\lambda)\Theta(C|\tau;\lambda),
\label{eq:jointdistribution}
\end{equation}
where $\psi(\tau;\lambda)$ is the probability density of the lifespan
$\tau$ and $\Theta(C|\tau;\lambda)$ is the probability that the coverage
is $C$, given that the lifespan is $\tau$.  The usual scaling assumption
for $\psi(\tau;\lambda)$, near the critical point, is
\cite{Marrobook}
\begin{equation}
\psi(\tau;\lambda) = \tau^{-1-\delta}\hat{f}\left[(\lambda_c - \lambda)\tau^\sigma \right],
\label{eq:probabilitydensity}
\end{equation}
for $\tau> \tau_{min}$, some minimum time scale. The scaling function
$\hat{f}(x)$ is non-symmetric, continuous at $x=0$, constant when
$|x| \ll 1$, and decays faster than a power law when $|x| \gg 1$. The
scaling hypothesis Eq.~(\ref{eq:probabilitydensity}) can be used to
derive a scaling relation between the exponent $\beta$ and the exponents
$\delta$ and $\sigma$. Below the critical point, $\lambda< \lambda_c$,
all realizations are finite and, thus,
$\int \psi(\tau;\lambda) d\tau=1$. Above this point, there is a finite
probability that a realization is endemic and, therefore,
$\int \psi(\tau;\lambda) d\tau=1-\mathrm{P_{end}}(\lambda)$. Combining
these two results and Eq.~(\ref{eq:probabilitydensity}) leads to
\begin{equation}
\mathrm{P_{end}}(\lambda) \sim \frac{1}{\sigma} (\lambda-\lambda_c)^{\frac{\delta}{\sigma}} \int_{0}^{\infty} x^{-1-\frac{\delta}{\sigma}}\left[\hat{f}(x)-\hat{f}(-x)\right] dx,
\label{P_end}
\end{equation}
which provides the relation $\beta=\delta/\sigma$. 

The scaling assumption
Eq.~(\ref{eq:probabilitydensity}) tells us that the lifespan is power
law distributed up to the cutoff value
\begin{equation}
  \tau_{\mathrm{cut}} \sim |\lambda_c-\lambda|^{-1/\sigma},
\label{tau_c}
\end{equation}
depending on the deviation from the critical point\footnote{Note that
  the pre-factor in Eq.~\eqref{tau_c} can be different when approaching
  the critical point from below or from above.}. In turn, this implies
that, close to the critical point, the moments $\langle\tau^n\rangle$
behave as~\footnote{A more precise calculation using the scaling 
assumption Eq.~\eqref{eq:probabilitydensity} is given by
\[
\langle\tau^n\rangle=\frac{a_{\pm}}{\sigma} |\lambda_c-\lambda|^{\frac{\delta-n}{\sigma}} \; \mbox{with} \; a_{\pm}=\int_0^{\infty} x^{\frac{n-\delta}{\sigma}-1 }f(\mp x)dx
\]
where the positive (negative) value means approaching the critical point from the right (left).
}
\begin{equation}
  \langle\tau^n\rangle \approx \int_0^{\tau_{\mathrm{cut}}} \tau^{n-1-\delta} \;
  d\tau \sim
  |\lambda_c - \lambda|^{\frac{\delta-n}{\sigma}}. 
\label{eq:moments}
\end{equation}
This result is similar to the behavior of the size of finite clusters in
regular percolation~\cite{stauffer94}.

\begin{figure}[t]
\centering
\includegraphics[width=8cm]{\FigPath/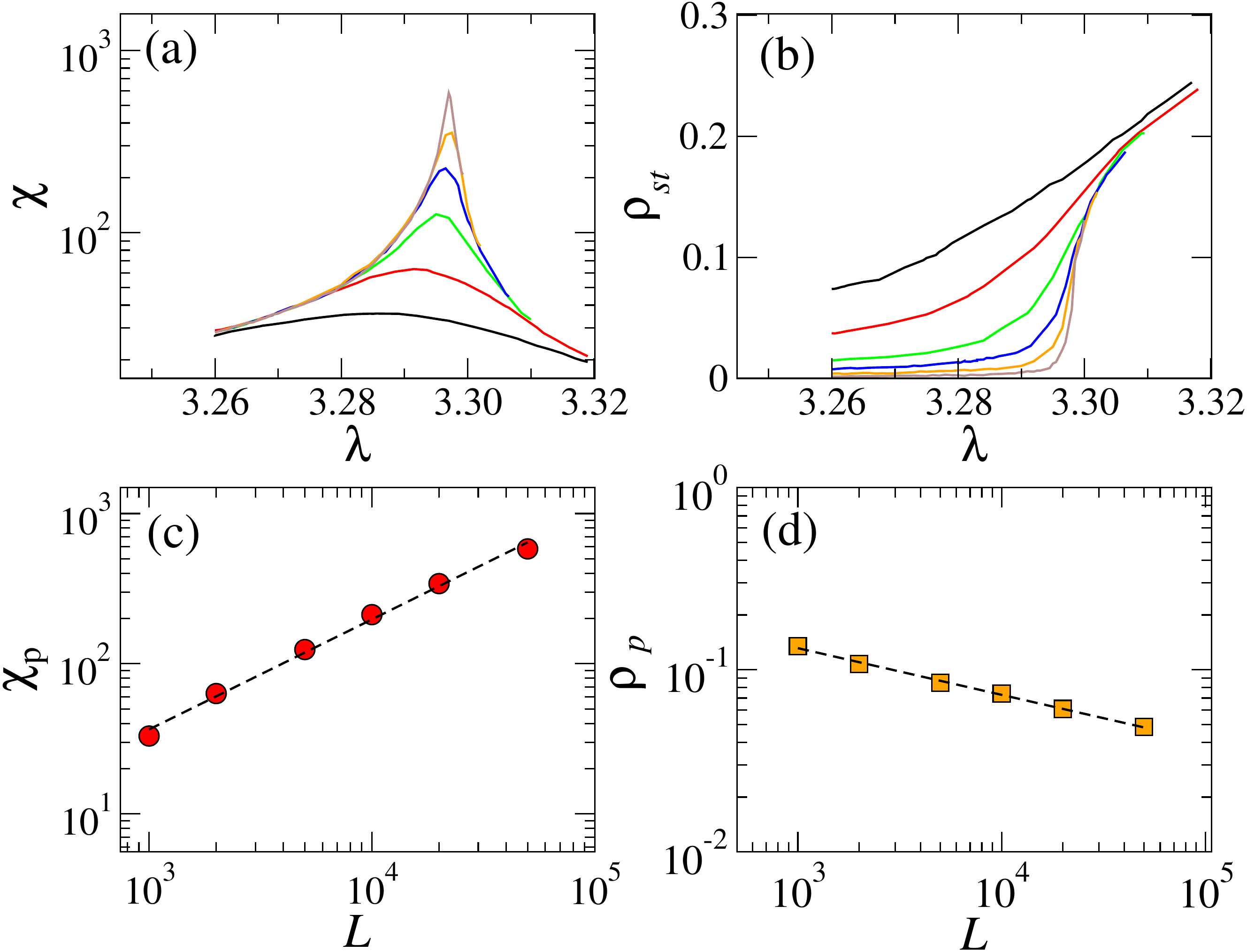}
\caption{(color online) Susceptibility (a) and density of active nodes
  (b) as a function of $\lambda$ for quasi-stationary simulations of
  the CP on 1$d$ lattices of different sizes. System size is $L=1000, 2000,
5000, 10000, 20000$, and $50000$ bottom to top in (a) and top to bottom in (b).
  Bottom plots show the
  size dependence of the height of the peak of the susceptibility,
  $\chi_p(L)$ (c), and the quasi-stationary density evaluated at the
  peak of the susceptibility, $\rho_{p}(L)$ (d). Dashed lines are
  power law fits of exponents $\beta/\nu_{\bot} = 0.253(5)$ and
  $(\gamma+\beta)/\nu_{\bot} =0.736(3)$. We performed $200$ individual
  realizations and $t = 2 \times 10^6$ Monte Carlo time steps. Error
  bars for $\chi_p(L)$ and $\rho_p(L)$ are smaller than symbols.}
 \label{fig:SUS}
\end{figure}

In finite systems, an additional temporal cutoff competes with
$\tau_{\mathrm{cut}}$ in Eq.~\eqref{tau_c}, namely, the temporal cutoff
$\bar{\tau}_\mathrm{cut}(N)$ arising from the finiteness of the system size. To
define this temporal scale, we consider the behavior near the critical
point of the average coverage
$\bar{C}(\tau;\lambda_c)\equiv\sum_C C\Theta(C|\tau;\lambda_c)$. At the critical 
point, we expect all physical observables to satisfy scaling relations. Thus, we can write
\begin{equation}
\bar{C}(\tau;\lambda_c) \sim \tau^\mu.
\label{eq:coverage}
\end{equation}
However, since $\bar{C}(\tau;\lambda_c)$ cannot become larger than $N$,
Eq.~(\ref{eq:coverage}) can only hold up to a cut-off value
$\bar{\tau}_{\mathrm{cut}}(N) \sim N^{1/\mu}$~\footnote{Notice that the fluctuations of the coverage near its maximum value $C \lesssim N$
  vanishes and, thus, in this region Eq.~\eqref{eq:coverage} can be considered as a
  deterministic equivalence between coverage and lifespan.}.  The
interplay between the two cut-offs present in the system,
$\tau_{\mathrm{cut}} \sim |\lambda_c - \lambda|^{-1/\sigma}$ (due to the
distance from the critical point) and $\bar{\tau}_{\mathrm{cut}}(N)$
(due to the finite size) determines the scaling of the moments
$\langle \tau^n \rangle$.  When
$\bar{\tau}_{\mathrm{cut}}(N) \gg \tau_{\mathrm{cut}}$, the system does
not notice its finiteness and, therefore, all moments are given by
Eq.~\eqref{eq:moments}.  Instead, when
$\bar{\tau}_{\mathrm{cut}}(N) \ll \tau_{\mathrm{cut}}$, the distribution
is cut-off by $\bar{\tau}_{\mathrm{cut}}(N)$ and, thus all moments
behave as
$\langle\tau^n\rangle \sim \int^{\bar{\tau}_{\mathrm{cut}}(N)}
\tau^{n-1-\delta}d \tau \sim [\bar{\tau}_{\mathrm{cut}}(N)]^{n-\delta}$.
To sum up:
\begin{equation}
 \langle\tau^n\rangle \sim \left\lbrace \begin{array}{lll}
       |\lambda_c-\lambda|^{\frac{\delta-n}{\sigma}} & ~~ & \text{if }~~|\lambda_c-\lambda|N^{\sigma/\mu} \gg 1  \\
       N^{\frac{n-\delta}{\mu}} & ~~ & \text{if }~~|\lambda_c-\lambda|N^{\sigma/\mu} \ll 1 
       \end{array}\right..
\label{eq:moments2}
\end{equation}
Defining the exponents $\gamma_n \equiv (n-\delta)/\sigma$ and
$\nu_{\bot} \equiv \mu/\sigma$, the behavior of Eq.~\eqref{eq:moments2} can be captured by the
following finite size scaling form
\begin{equation}
\langle\tau^n(N)\rangle = N^{\gamma_n/\nu_{\bot}}G_n\left[(\lambda_c-\lambda)N^{1/\nu_{\bot}}\right],
\label{eq:fss_tau}
\end{equation}
where the scaling function $G_n(x)$ is constant if $|x| \ll 1$ and goes as
$|x|^{-\gamma_n}$ when $|x| \gg 1$. 
As usual, we expect to find a maximum of $\langle\tau^n(N)\rangle$ 
around a value $\lambda_p^{LS}(N)$, which depends on the system size as 
\begin{equation}
  \label{eq:1}
  \lambda_p^{LS}(N) = \lambda_c + A_{LS} N^{-1/\nu_{\bot}}.
\end{equation}
We can then use, in general, the average
lifespan to determine numerically the critical point and some of the
critical exponents. There is, however, a pathological case if the exponent of
the lifespan distribution is exactly $\delta=1$. 
In such a case, the average lifespan $\av{\tau}$ does diverge, but 
logarithmically; the critical point can still be determined but
critical exponents cannot. 
This problem disappears if one uses the second moment $\av{\tau^2}$
instead.
\begin{figure}[t]
\centering
\includegraphics[width=8cm]{\FigPath/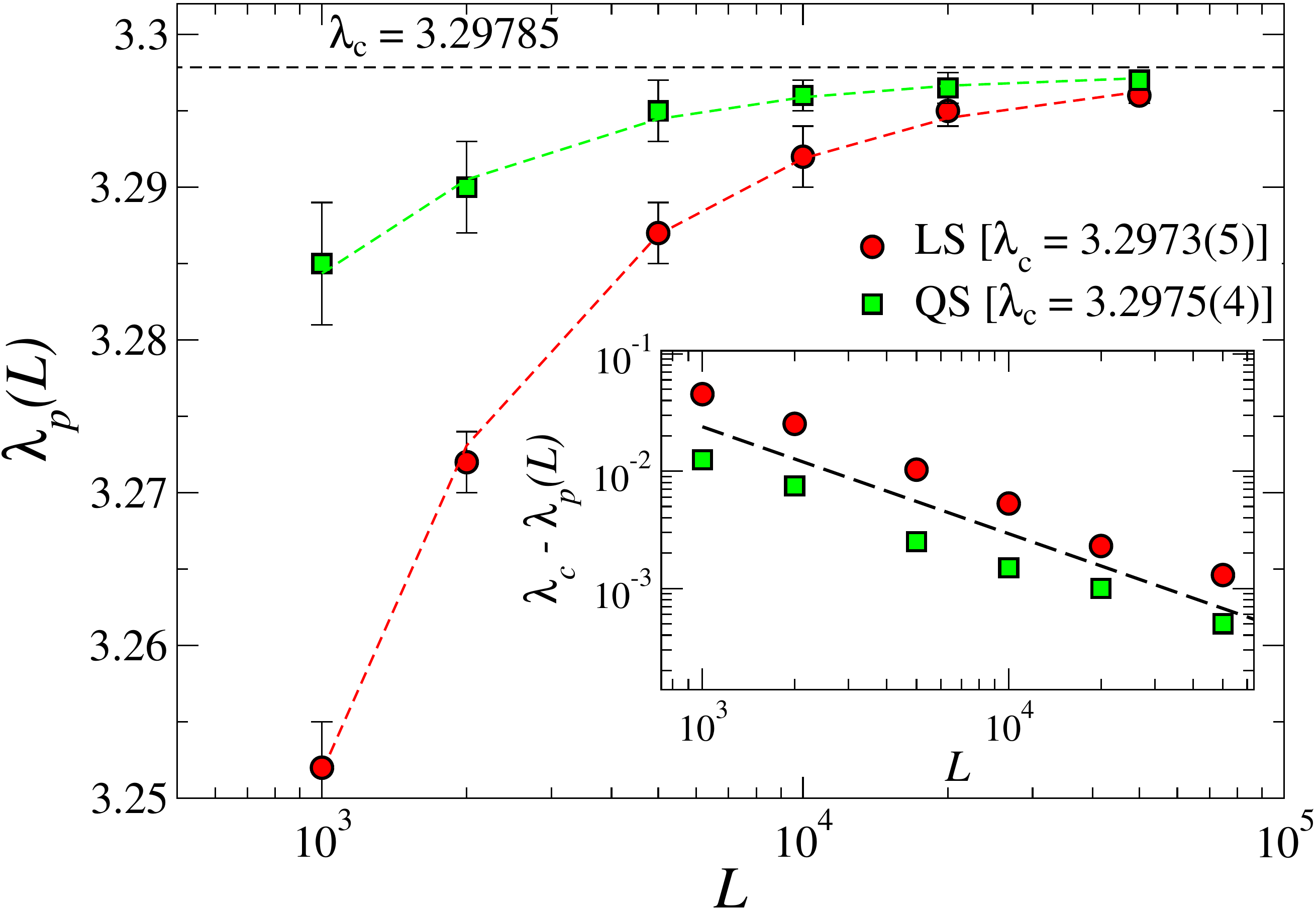}
\caption{(color online) Size dependence of the $\lambda_p(L)$ estimates
  of the transition point for the quasi-stationary and lifespan
  methods. Dashed lines are non-linear regressions used to determine the
  critical point and to estimate the critical exponent $\nu_{\bot}$,
  applying Eq.~(\ref{eq:3}), see Table~\ref{tab:values}. The horizontal
  line marks the accepted best estimate value of the critical
  point. Inset shows $\lambda_c-\lambda_p$ as a
    function of $L$ in log-log scale. 
    The dashed line has slope $1/\nu_\perp$ as a guide to the eyes. }
 \label{fig:nonlinear}
\end{figure}

Finally, concerning the order parameter  $\mathrm{P_{end}}(\lambda,N)$,
defined as the probability that a run is endemic, it fulfills the standard
finite-size scaling form 
\begin{equation}
 \mathrm{P_{end}}(\lambda,N) \varpropto N^{-\beta/\nu_{\bot}}
  f\left[(\lambda-\lambda_c)N^{1/\nu_{\bot}}\right]. 
\label{eq:prob}
\end{equation}
From this expression, we can determine the exponent $\beta/\nu_{\bot}$ by examining
the $N-$dependence of $\mathrm{P_{end}}(\lambda,N)$ at the critical point
\begin{equation}
  \mathrm{P_{end}}(\lambda_c,N) \varpropto N^{-\beta/\nu_{\bot}}.
  \label{eq:prob2}
\end{equation}
The missing piece of the scaling theory presented above is the value of
the exponent $\mu$, governing the scaling with the system size 
of the lifespan cut-off at criticality $\bar{\tau}_{\mathrm{cut}}(N)$.  
As we will check
numerically below, this cut-off can be identified in regular lattices
with the characteristic relaxation time, which close to criticality
scales as $\bar{\tau} \sim \vert \lambda - \lambda_c
\vert^{-\nu_\parallel}$ \cite{Marrobook}. Comparing this relation with Eq.~\eqref{tau_c} leads to the identity
$\mu=\nu_{\perp}/\nu_{\parallel}$.

\section{Numerical results}
\label{sec:1d}

The critical properties of the CP on a one-dimensional lattice and the
corresponding finite-size scaling theory for the transition are very
well known, and accurate theoretical and numerical values are readily
available for comparison \cite{Marrobook}. This makes the CP on a
one-dimensional lattice the ideal testbed for numerical methods. In this
section, we present results of numerical simulations of the CP on a
$d=1$ lattice, applying both the QS and LS methods. Hereafter, we use
$N=L$.

\subsection{Quasi-stationary simulations}
\begin{figure}[t]
\centering
\includegraphics[clip=true,trim=0cm 6cm 0cm 0cm, width=8cm]{\FigPath/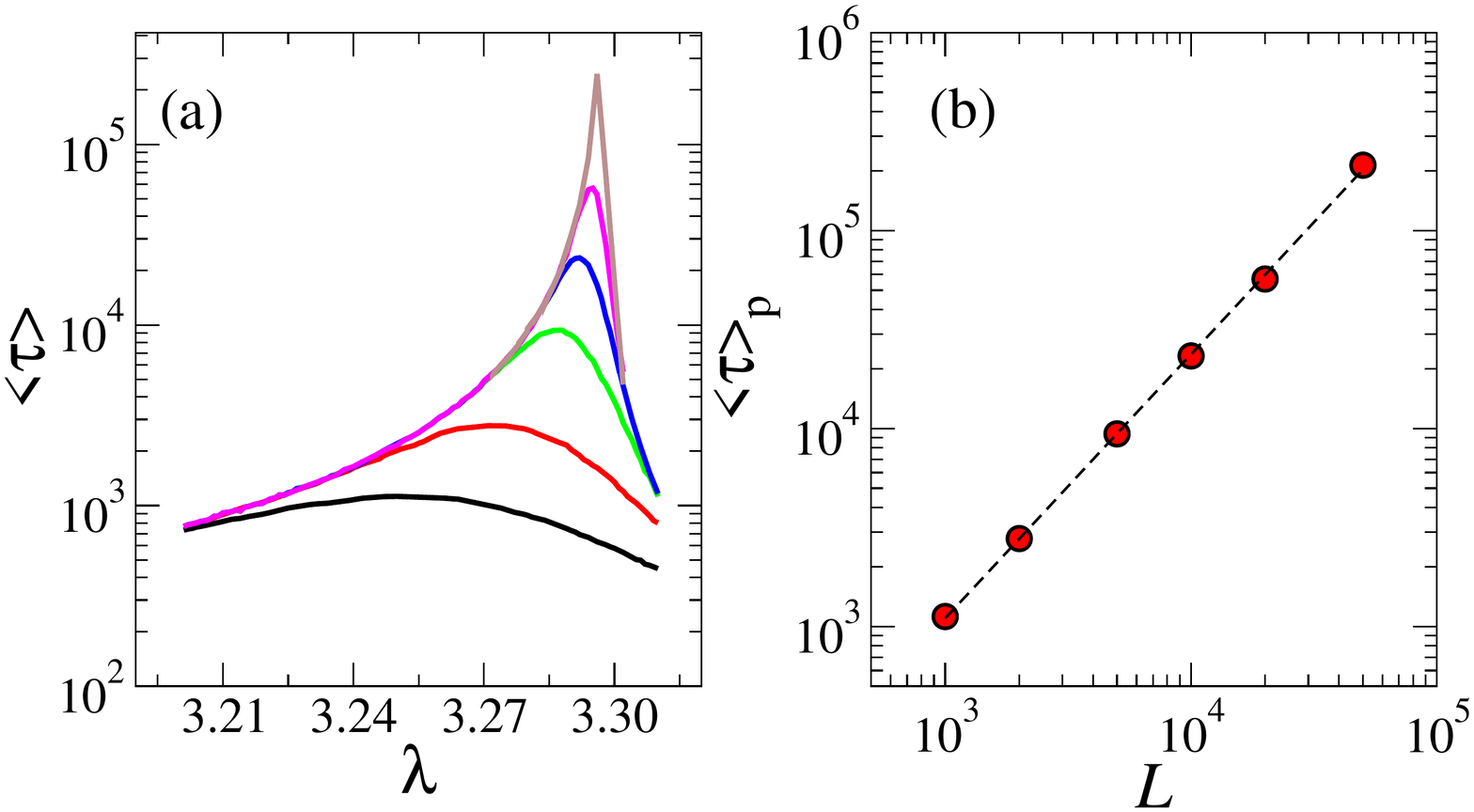}
\includegraphics[clip=true,trim=0cm 6cm 0cm 0cm, width=8cm]{\FigPath/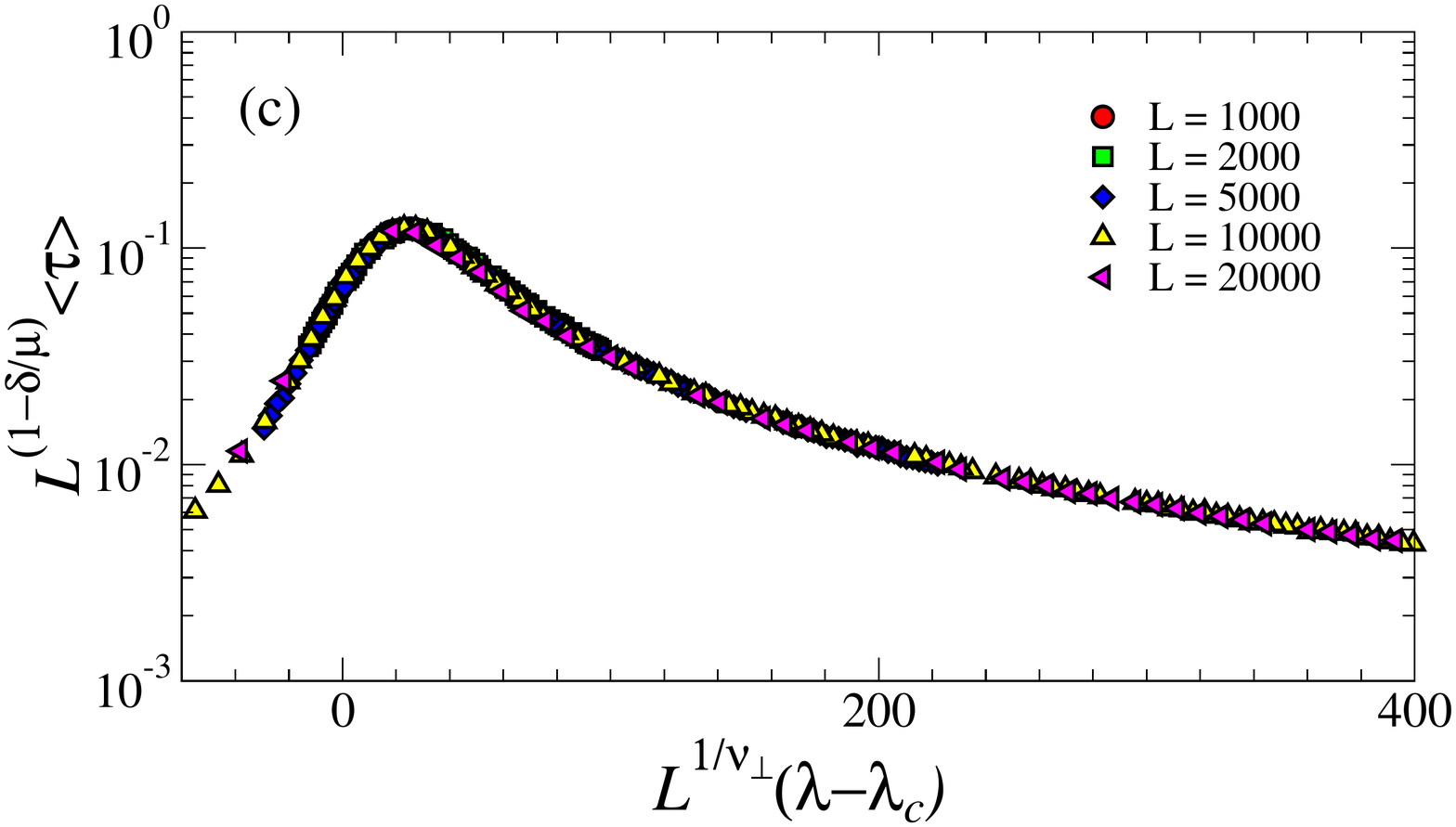}
\caption{(color online) (a) Lifetime $\av{\tau}$ against creation rate
  for the CP on a one dimensional lattice.  Curves are for system size
  (bottom to top) $L=1000, 2000, 5000, 10000, 20000$ and $50000$.  (b)
  Size dependence of the height of the peak of the average lifespan
  $\tau_p(L)$.  The dashed line represents a power law regression with
  slope $(1-\delta)/\mu = 1.32(1)$. We perform $5 \times 10^5$
  individual realizations for each size.  Error bars are smaller than
  symbols. (c) The scaling plot of the lifetime according to
  Eq.~\eqref{eq:fss_tau} for the same data of panel (a)}
 \label{fig:LS}
\end{figure}

In Fig~\ref{fig:SUS}, we show the results obtained by performing QS
simulations of the CP on a one-dimensional lattice of length $L$.  The
susceptibility $\chi$, Fig~\ref{fig:SUS}(a), shows a well defined peak,
which becomes narrower and taller as the system size $L$ grows.  The
plot of the quasi-stationary density $\rho_{st}(L)$,
Fig~\ref{fig:SUS}(b), also displays a transition becoming narrower and
sharper as $L$ grows. From the position of the susceptibility peak
$\lambda_p^{QS}(N)$ it is possible to obtain asymptotically an estimate
of the transition point $\lambda_c$ by applying the relation in
Eq.~(\ref{eq:3}). We have used this expression to perform a nonlinear
regression to determine the critical point $\lambda_c$ and the exponent
$\nu_{\bot}$, see Fig.~\ref{fig:nonlinear} and Table~\ref{tab:values}.
The values obtained by this procedure are in very good agreement with
the best estimates accepted in the
literature~\cite{Henkel}.

Right at the critical point, the average density of particles and the
susceptibility should scale with the system size as given by
Eq.~(\ref{eq:FFS}). From this analysis, see
Fig~\ref{fig:SUS}(c) and (d), we can compute the exponents
$\beta/\nu_{\bot}$ and $(\gamma+\beta)/\nu_{\bot}$, which again reproduce
with good accuracy the known values of the CP, see
Table~\ref{tab:values}. 
\begin{table}[b]
  \begin{ruledtabular}
    \begin{tabular}{|c|c|c|c|}
                       & Theoretical & QS          & LS        \\ 
    \hline
    $\lambda_c$         & 3.297848(22)  & 3.2975(4)   & 3.2973(5)  \\ 
    \hline \hline
    $\nu_\perp$          & 1.096854(4)  & 1.098(5)    & 1.100(5)   \\
    \hline
    $\beta / \nu_\perp$  & 0.252068(8)  & 0.253(5)    &   0.255(5)   \\
    \hline
    $(\gamma + \beta)/ \nu_\perp$ & 0.74792(2)  & 0.736(3)  &    ---     \\ 
    \hline
     $\mu$ ($=\nu_\perp / \nu_\parallel$) &0.632613(4)   &  ---  & 0.64(1)       
  \end{tabular}
  \end{ruledtabular}
  \caption{Critical point and exponents of the CP in a $d=1$ lattice
    obtained using the QS and LS methods. For comparison, we quote
    also the best estimates of those, from
    Ref.~\cite{Henkel}.}
  \label{tab:values}
\end{table}

\subsection{The lifespan method}

As discussed in Sec.~\ref{sec:fss}, in the LS method for finite systems, 
the role of the order parameter is played by the probability 
$\mathrm{P_{end}}(\lambda,L)$ that a run reaches
the predefined coverage $\Ct$ (i.e. it is effectively endemic), while
the analogue of the susceptibility is given by the average duration
$\av{\tau}$ of finite realizations.
\begin{figure}[t]
\centering
\includegraphics[width=8cm]{\FigPath/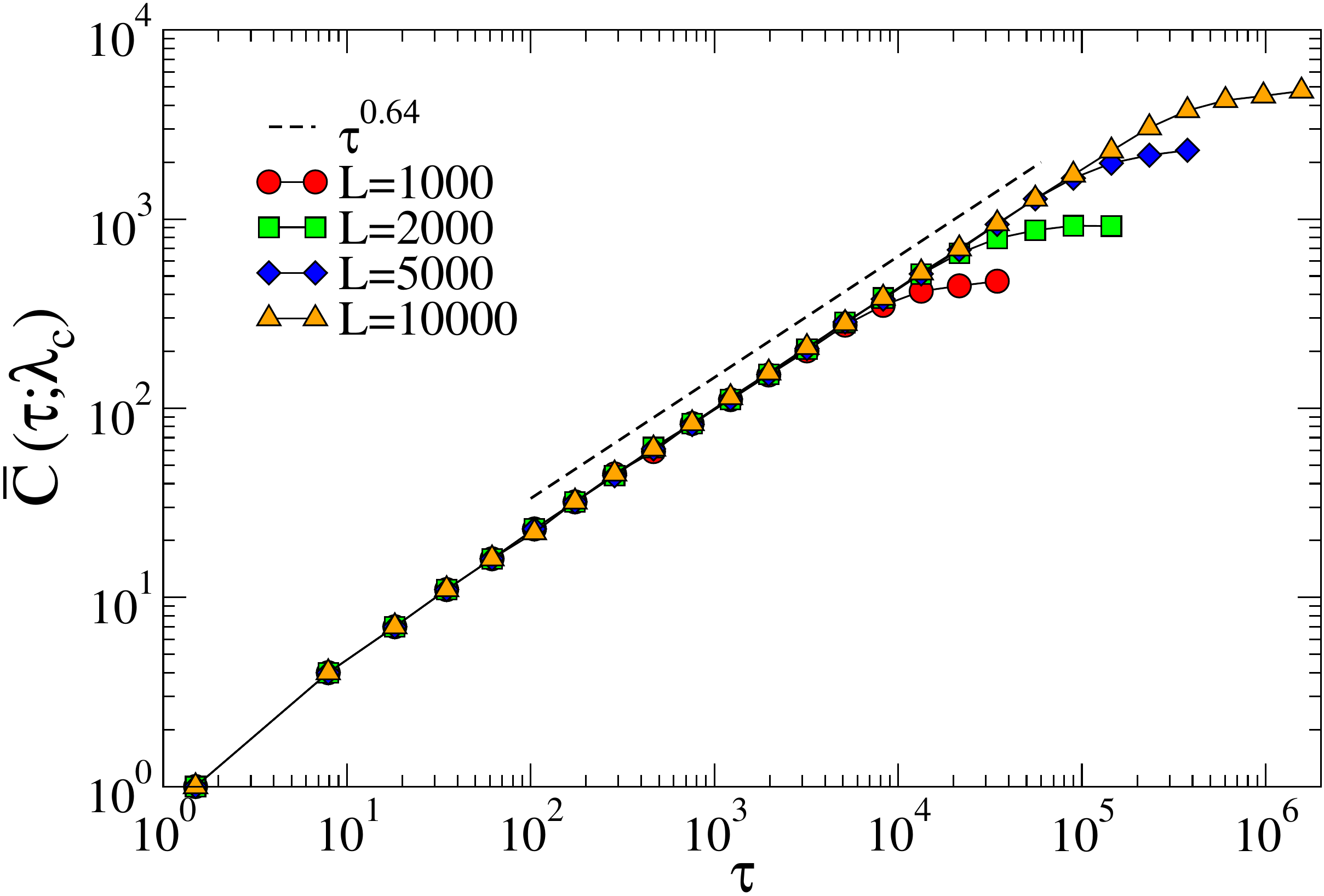}
\caption{(color online) Average coverage as a function of $\tau$
  evaluated at $\lambda = \lambda_p^{LS}(N)$.  The dashed line has
  slope $0.64$ and serves as a guide to the eyes.}
 \label{fig:coverage}
\end{figure}

\begin{figure}[t]
\centering
\includegraphics[width=8cm]{\FigPath/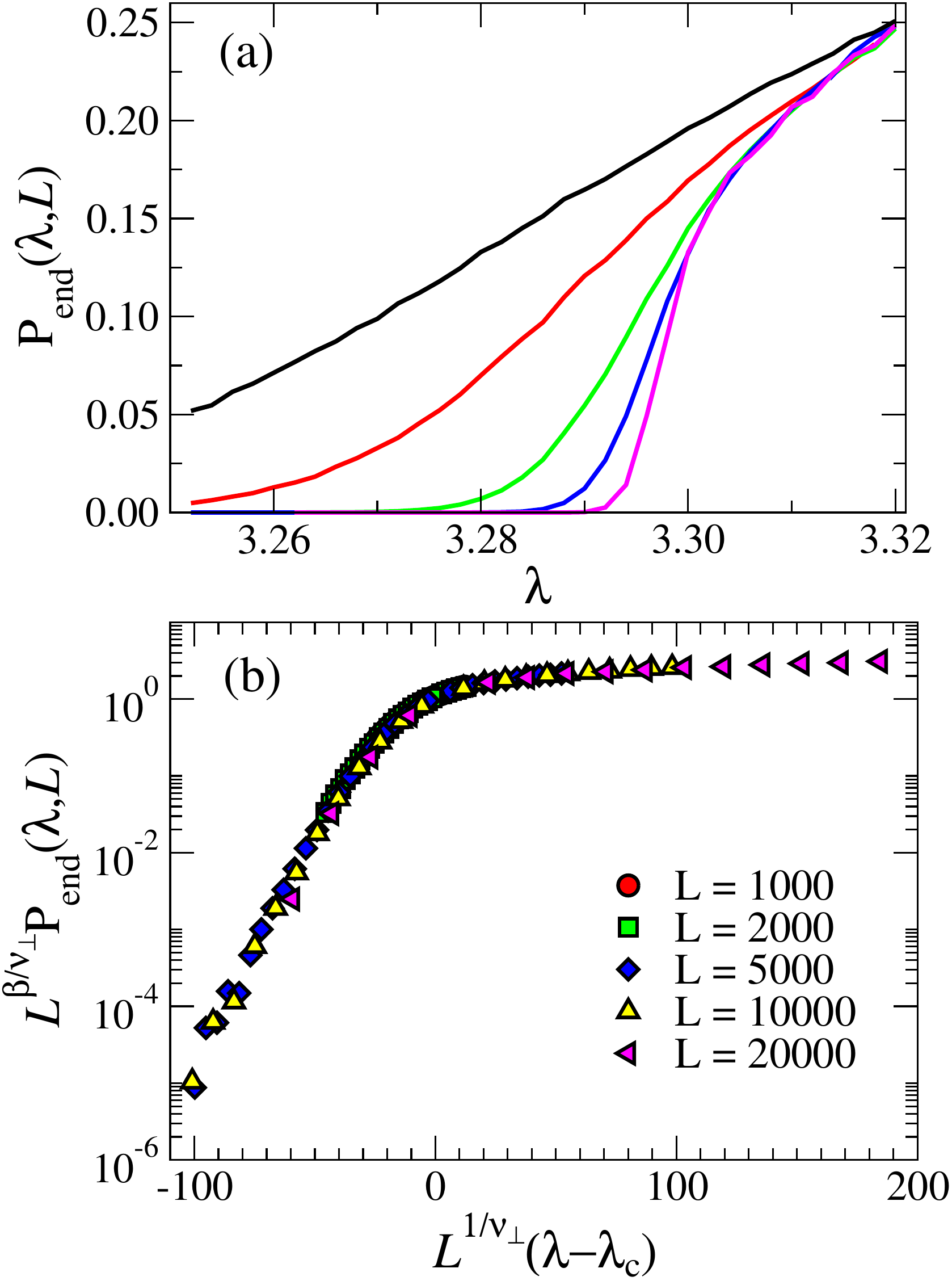}
\caption{(color online) (a) Probability to reach the predefined coverage
  fraction $\ct=0.5$ as a function of $\lambda$ for the CP on a $1-d$
  lattice.  Curves are for system size (top to bottom) $L=1000, 2000,
  5000, 10000, 20000$. In (b),
  we show the scaling plot of this probability according to
  Eq.~\eqref{eq:prob}.}
 \label{fig:prob}
\end{figure}
In Fig.~\ref{fig:LS}(a), we plot the average lifespan $\av{\tau}$ as a
function of $\lambda$, for different system sizes, computed for a fixed
coverage fraction threshold $\ct = 0.5$; the effect of varying the coverage
fraction threshold is discussed in Sec.~\ref{sec:appendixA}. 
From this figure, we
can observe that the lifespan $\av{\tau}$ has a well-defined peak at a
value $\lambda_p^{LS}(L)$, signaling the presence of a phase transition. The
dependence of the peak position as a function of the system size $L$ is
reported in Fig.~\ref{fig:nonlinear}.  A non-linear fitting of the data
according to Eq.~(\ref{eq:3}) provides numerical estimates for the
critical point $\lambda_c$ and the exponent $\nu_{\bot}$, see
Table~\ref{tab:values}, which are compatible with the exact results
derived analytically. Hence, we conclude that both the QS and the LS
method recover compatible results for the position of the critical point and the exponent $\nu_{\bot}$.  

The peak value $\av{\tau}_p$ of the average lifespan grows as a
power-law as a function of $L$, see Fig.\ref{fig:LS}(b).  According to
the scaling theory presented in Sec.~\ref{sec:fss}, the exponent of this
growth is equal to $(1-\delta)/\mu$, for which we obtain a value
$1.32(1)$.  The value of $\delta$ is well-known in the literature,
namely $\delta =0.159464(6)$ \cite{Henkel}. From here, we obtain the
exponent $\mu = 0.64(1)$. We can also determine this exponent directly
from the scaling of the average coverage near the critical point,
$\bar{C}(\tau;\lambda_c) \sim \tau^\mu$, see Eq.~\eqref{eq:coverage}. In
Fig.~\ref{fig:coverage}, we analyze this coverage, obtaining numerically
an exponent $\mu = 0.64(1)$, in perfect agreement with the value
found from the scaling of the peak of the average lifespan.  In
Fig.~\ref{fig:LS}(c), we finally check the full finite-size scaling form
of the lifespan $\av{\tau}$ as given by Eq.~(\ref{eq:rho2}). We perform
a data collapse analysis by plotting $L^{(1-\delta)/\mu} \av{\tau}$ as a
function of $L^{1/\nu_\perp} ( \lambda - \lambda_c )$.  The perfect
collapse of the plots shown in Fig.~\ref{fig:LS}(c) confirms the
validity of the finite-size scaling proposed in Eq.~(\ref{eq:rho2}).

Concerning the order parameter, in Fig.~\ref{fig:prob}(a) we plot
$\mathrm{P_{end}}(\lambda,L)$ evaluated with threshold coverage fraction
$\ct=0.5$ as a function of $\lambda$ and different values of $L$. As we
can see, it displays a sharp phase transition at the critical point
when the size of the system increases. Close to criticality, and for
large $L$, this probability exhibits a power law form with system size
given by Eq.~(\ref{eq:prob2}). By analyzing
$\mathrm{P_{end}}(\lambda_c,L)$ as a function of $L$, we can obtain the
exponent $\beta/\nu_{\bot}$, see Table~\ref{tab:values}, again in very
good agreement with QS estimates. Finally, in Fig.~\ref{fig:prob}(b), we
check the full finite size scaling form Eq.~(\ref{eq:prob}) by plotting
$L^{\beta/\nu_\perp } \mathrm{P_{end}}(\lambda,L)$ as a function of
$L^{1/\nu_\perp } ( \lambda -\lambda_c )$, using the numerical exponents
found. The perfect data collapse found demonstrates, once again, the
correctness of the finite-size scaling form for the order parameter of
the LS method.

\subsection{Robustness with respect to the coverage fraction threshold $\ct$}
\label{sec:appendixA}
\begin{figure}[t]
\centering
\includegraphics[width=8cm]{\FigPath/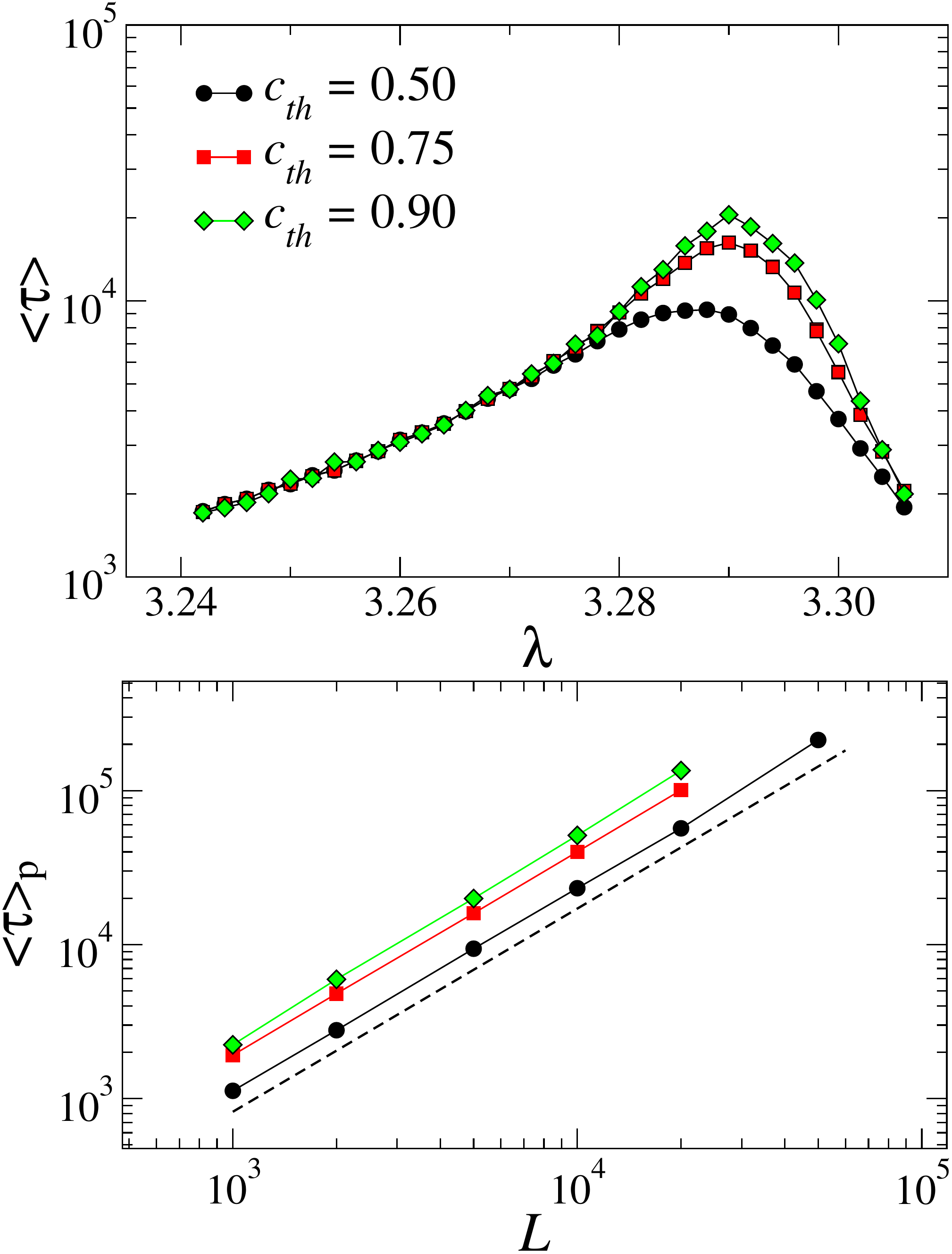}
\caption{(color online)(a) The average lifespan
  $\av{\tau}$ against creation rate for different values of $\ct$, for
  $L = 5000$. (b) The height of the peak grows with an exponent
  independent of $\ct$. The dashed line has slope $1.32$.}
 \label{fig:coverage_appendix}
\end{figure}

In the results presented above, we have used a fixed value of the
coverage fraction threshold $\ct=0.5$. As we have discussed in Sec. IV, our
results are however independent of the precise value of $\ct$. 
To check such a claim, we perform additional
simulations for threshold values $\ct=0.75$ and $\ct=0.90$. In
Fig.~\ref{fig:coverage_appendix}(a), we plot the average lifespan as a
function of $\lambda$ for a fixed system size, $L=5000$, and different
values of the coverage fraction threshold $\ct$. As we can see, increasing the
coverage fraction threshold slightly shifts both the position of the peak as
well as the height of the maximum lifespan. Nevertheless, as we show
in Fig.~\ref{fig:coverage_appendix}(b) the height of the peak of
$\av{\tau}$ scales with the system size $L$ with an exponent
$(1-\delta)/\mu = 1.32$, that is independent of $\ct$.

\begin{figure}[t]
\centering
\includegraphics[width=8cm]{\FigPath/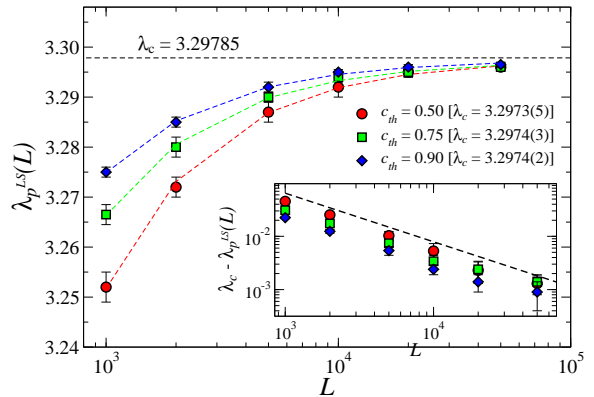}
\caption{(color online) Size dependence of $\lambda_p(L)$ estimates
  with lifespan simulation method with different values of $\ct$. 
  Inset shows $\lambda_c-\lambda_p$ as a
    function of $L$ in log-log scale. The dashed line has slope $1/\nu_\perp$ as a guide to
    the eyes. }
 \label{fig:coverage2_appendix}
\end{figure}

In Fig.~\ref{fig:coverage2_appendix}, we show the extrapolation of the
different values of the position of the peak $\lambda_p^{LS}(L)$,
applying Eq.~(\ref{eq:1}).  As we can see, all values of $\ct$ lead
asymptotically to the same value of $\lambda_c$. This fact indicates
that the critical properties of the model are recovered the LS method in
a robust way, independently of the arbitrary choice of the coverage fraction
threshold $\ct$. In this sense, it is noticeable that, although the
$\lambda_p^{QS}(L)$ values obtained via the QS method approach the
critical point faster than those obtained using LS, see
Fig.~\ref{fig:nonlinear}, if the value of the coverage fraction 
threshold $\ct$ in the LS simulation is increased, the convergence 
to the asymptotic value of the critical point becomes faster in $L$, although
computationally more expensive.

\section{Conclusions}
\label{sec:conclusions}

The precise determination of the critical properties of 
absorbing-state phase transitions is a crucial problem in
non-equilibrium statistical mechanics. Indeed, while powerful analytical
strategies, such as field theoretic methods and their renormalization
group analysis, are available, these methods are technically complex and
ensuing loop expansions lead to approximate values for critical
exponents, sometimes of uncontrolled validity in physical
dimensions. For this reason, good numerical tools are of invaluable
help.  Here, we have reported a new numerical technique, the lifespan
method, which is able to determine with great accuracy the critical
properties of absorbing-state phase transitions.  To this end, we have
developed the corresponding finite-size scaling theory, which allows us
to determine precisely both the the critical point and the critical
exponents by looking at the size dependence of the associated
susceptibility and order parameter. Results of the application of the
lifespan method to the contact process in a $d=1$ lattice are compared
with results from the quasi-stationary method and other numerical and
analytical results, showing that the new approach is fully reliable. We
note that, even though the LS method has been validated here for an
absorbing-state phase transition to a unique absorbing state, it can be
generalized to systems with many such states.

To sum up, the lifespan method is an alternative way to numerically
studying systems with absorbing states, which complements more
traditional techniques, such as the quasi-stationary method, and that
will represent in the future a useful addition to the numerical toolset
of the statistical physics practitioner.

\begin{acknowledgments}
  R.P.-S. acknowledges financial support from the Spanish MINECO, under
  projects No. FIS2010-21781-C02-01 and FIS2013-47282-C2-2, EC
  FET-Proactive Project MULTIPLEX (Grant No. 317532), and ICREA
  Academia, funded by the Generalitat de Catalunya.. R.P.-S. and
  A.S.M. acknowledge financial support from CAPES under project
  No.5511-13-5. M.~B. acknowledges financial support from the James
  S. McDonnell Foundation; the ICREA Academia foundation, funded by the
  {\it Generalitat de Catalunya}; MINECO projects No.\
  FIS2010-21781-C02-02 and FIS2013-47282-C2-1-P; and {\it Generalitat de
    Catalunya} grant No.\ 2014SGR608.

\end{acknowledgments}


%

\end{document}